\documentclass[cits]{JINST}
\usepackage{graphicx}
\usepackage{epsfig}
\usepackage{subfig}
\newcommand{\subfigure}{\subfloat}
\title{Investigation of a direction sensitive sapphire detector stack at the $5$ $GeV$ electron beam at DESY-II}
\author{O.~Karacheban$^a$$^b$, K.~Afanaciev$^c$, M.~Hempel$^a$$^b$, H.~Henschel$^a$, W.~Lange$^a$,
 J.L.~Leonard$^a$, I.~Levy$^e$, W.~Lohmann$^b$$^f$ and  S.~Schuwalow$^d$\\
\llap{$^a$}Deutsches Elektronen-Synchrotron, Zeuthen, Germany\\
\llap{$^b$}Brandenburgische Technische Universit{\"a}t  Cottbus, Cottbus, Germany\\
\llap{$^c$}NC PHEP BSU, Minsk, Belarus\\
\llap{$^d$}Deutsches Elektronen-Synchrotron, Hamburg, Germany\\
\llap{$^e$}Tel Aviv University, Tel Aviv, Israel\\
\llap{$^f$}CERN, Geneva, Switzerland\\

E-mail: \email{olena.karacheban@desy.de}}

\abstract{Extremely radiation hard sensors are needed in particle physics experiments to instrument the region near the beam pipe. 
Examples are beam halo and beam loss monitors at the Large Hadron Collider, FLASH or XFEL. 
Currently artificial diamond sensors are widely used. In this paper 
single crystal sapphire sensors are considered as a promising alternative. 
Industrially grown sapphire wafers are available in large sizes, are of low cost and, like diamond sensors, can be operated without cooling.
Here we present results of an irradiation study done with sapphire sensors in a high intensity low energy electron beam. Then, 
a multichannel direction-sensitive sapphire detector stack is described. 
It comprises 8 sapphire plates of 1~$\rm{cm^2}$ size and $525$~$\rm{\mu m}$ thickness, metallized on both sides, and apposed to form a stack. 
Each second metal layer is supplied with a bias voltage, and the layers in between are connected to charge-sensitive preamplifiers. 
The performance of the detector was studied in a $5$ GeV electron beam. 
The charge collection efficiency measured as a function of the bias voltage 
rises with the voltage, reaching about $10\% $ at $950$~V. 
The signal size obtained from electrons crossing the stack at this voltage is about $22000$ e, where e is the unit charge. 

The signal size is measured as a function of the hit position, showing variations of up to 20\% in the direction perpendicular to the beam and to the electric field.
The measurement of the signal size as a function of the coordinate parallel to the electric field 
confirms the prediction that mainly electrons contribute to the signal. 
Also evidence for the presence of a polarisation field was observed.}


\begin{document}

\section{Introduction}

For the operation in a harsh radiation environment, typical for near-beam detectors at LHC or free electron lasers like FLASH
and XFEL, extremely radiation hard sensors are needed. 
Currently CVD grown diamond sensors are applied e.g. for machine induced background and on-line luminosity measurements~\cite{bcm1f,flashbhm}. 
Regardless of the excellent radiation hardness and low leakage current at room temperature, the application of diamond sensors 
is limited due to high cost, relatively small size and low manufacturing rate. 
As an alternative we suggest to use sapphire sensors. 
Optical grade single crystal sapphire is industrially grown in practically unlimited amount and the wafers are of large size and low cost. 
Sapphire sensors have been used so far in cases where the signal is generated by simultaneous hits of many particles, i.e. 
in the beam halo measurement at FLASH, and are planned to be installed at FLASH~II, XFEL and the CMS experiment at the LHC. 
It was found that the time characteristics of signals from sapphire sensors are similar to the ones from CVD diamond sensors~\cite{flashbhm}. 
The radiation hardness of sapphire sensors was studied in a low energy electron beam
up to an absorbed dose of 12 MGy~\cite{CARAT}. The charge collection efficiency, CCE, as a function of the dose will be presented.
Furthermore, a detector composed of metallized sapphire plates of $10\times 10~\rm{mm^2}$ area and $525~\rm{{\mu}m}$ thickness   
to be used for single particle detection is investigated in a $5$~GeV electron beam. The total thickness of this detector amounts 
to 14\% of a radiation length. Since the response is depending on the direction of the particles crossing it, interesting fields of 
applications are beam-halo rate 
or low angle scattering measurements.  
Basic characteristics, like the dependence of the CCE on the applied voltage and position resolved sensor response, are reported 
and compared to a model of the charge transport in the presence of polarisation. 

\section{Basic features of artificial Sapphire}

Sapphire is a crystal of aluminum oxide, Al$_2$O$_3$.
The sapphire sensors were obtained from the CRYSTAL company~\cite{CRYSTAL}. Single crystal
ingots were produced using the Czochralski method and cut into wafers of $525~\rm{\mu m}$ thickness. Contamination
of other elements are on the level of a few ppm. Relevant properties of sapphire are listed in 
Table~\ref{tab:table1}. For comparison, the same properties are also given for diamond.  
\begin{table}[h!]
 \begin{center}
  \caption{Relevant material properties of single crystal artificial sapphire and diamond~\cite{CRYSTAL},~\cite{MolTech}.} 
  \label{tab:table1}
    \begin{tabular}{|l|c|c|}
      \hline
       Material properties                          &  sapphire   &   diamond               \\
      \hline
       density, g / cm$^3$                  &  3.98       &    3.52                     \\          
       bandgap, eV                        &  9.9        &    5.47                     \\
       energy to create an eh pair, eV    &  27         &    13                       \\ 
       dielectric constant (depending on the orientation) & 9.3 - 11.5  &    5.7                      \\
       dielectric strength, V / cm    &  4$\times$10$^5$         &    10$^6$                     \\
       resistivity, Ohm$\cdot$cm at 20$^{\circ}$ C  &  10$^{16}$ &  10$^{16}$                 \\
       electron mobility, cm$^2$ / (V$\cdot$s) at 20$^{\circ}$ C&  600 &  2800                  \\
       \hline  
\end{tabular}
   \end{center}
\end{table}
The band gap of sapphire is larger than for diamond, resulting in a by a factor
of two larger energy needed to create an electron-hole pair. The energy loss of a charged particle moving in sapphire is, 
however, larger than in diamond. The amount of charge carriers generated per unit length in sapphire is about 60 \% of
of the amount generated in diamond.

\section{The response of sapphire sensors as a function of the dose }

\begin{figure}[ht!]
  \begin{center}
  {\includegraphics[width=12.0cm]{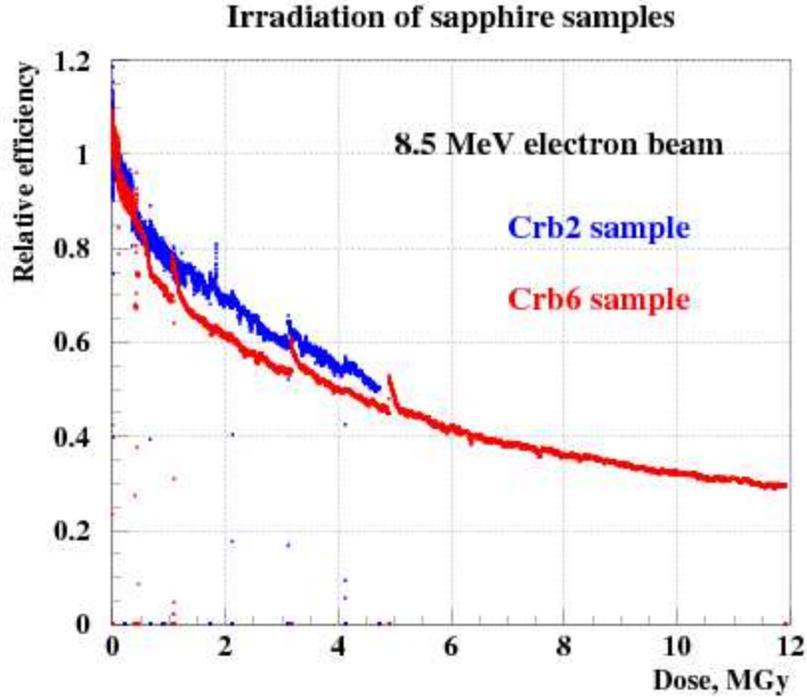}}
   \end{center}
\caption []{\label{fig:RadHard}
The relative CCE as a function of the dose in an electron beam for two sapphire sensors.
}
\end{figure}

Two sensors $10\times 10~\rm{mm^2}$ size and 525 $\rm{\mu}m$ thickness 
were exposed to a high-intensity electron beam at the linear accelerator DALINAC at TU Darmstadt, Germany. 
The beam energy was $8.5$~MeV, a typical value for electrons and positrons in the electromagnetic 
shower maximum for the near-beam calorimetry
at the future linear collider~\cite{pcvd-IEEE}. 
The response of the sensors was measured as the signal current. 
The relative drop of the signal current, interpreted as the relative drop of the charge collection efficiency, CCE,
is shown in Figure~\ref{fig:RadHard} for both sensor samples. 
As can be seen, the CCE degrades to about 30\% of the initial CCE after a dose of $12$~MGy, 
corresponding to more than 10 years of operation at
e.g. the ILC~\cite{ILC} at nominal beam parameters at $500$~GeV centre-of-mass energy~\cite{dose_estimate}.
The peaks on the rather smooth curves show an increase of the CCE by about 10 \% 
after periods when the beam was switched off to allow other intermediate
measurements or because of beam losses. These breaks were about a few minutes. Assuming that the decrease of response  
is partly caused by a reduced field inside the sensor due to polarisation, the short increase of the CCE indicates a release 
of trapped charge carriers with a corresponding decay constant. In addition, no current was seen during beam interruptions, indicating that 
the dominant trap release mechanism is recombination.  
When the beam was switched back on, the CCE continued to decrease. 
The leakage current of the sensors was measured before and after irradiation to be below 10 pA. 
      
\section{Detector stack design}

\begin{figure}[ht!]
  \begin{center}
   \subfigure[]
  {\includegraphics[width=7.0cm]{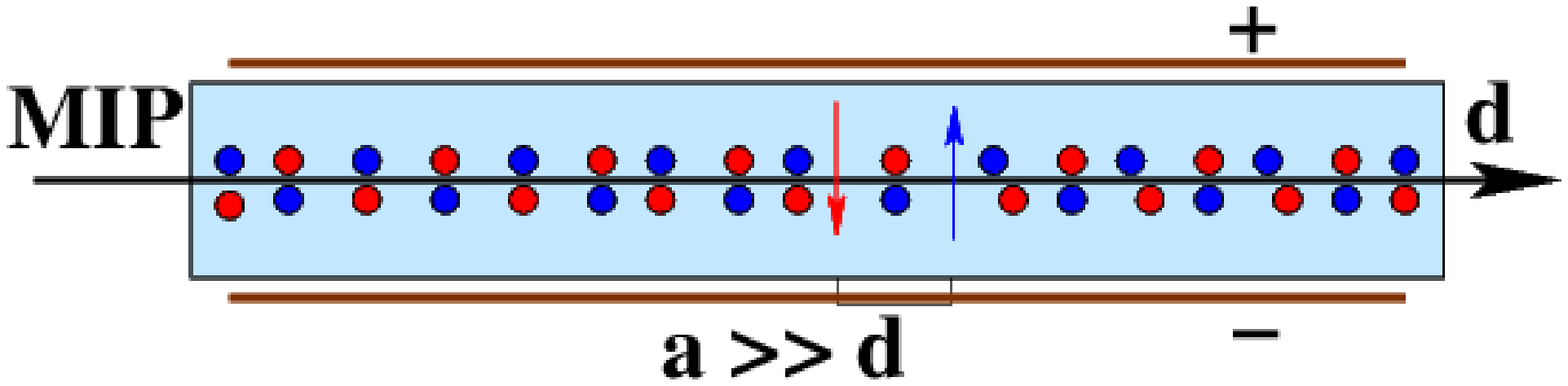}}
    \subfigure[]
  {\includegraphics[width=8.0cm]{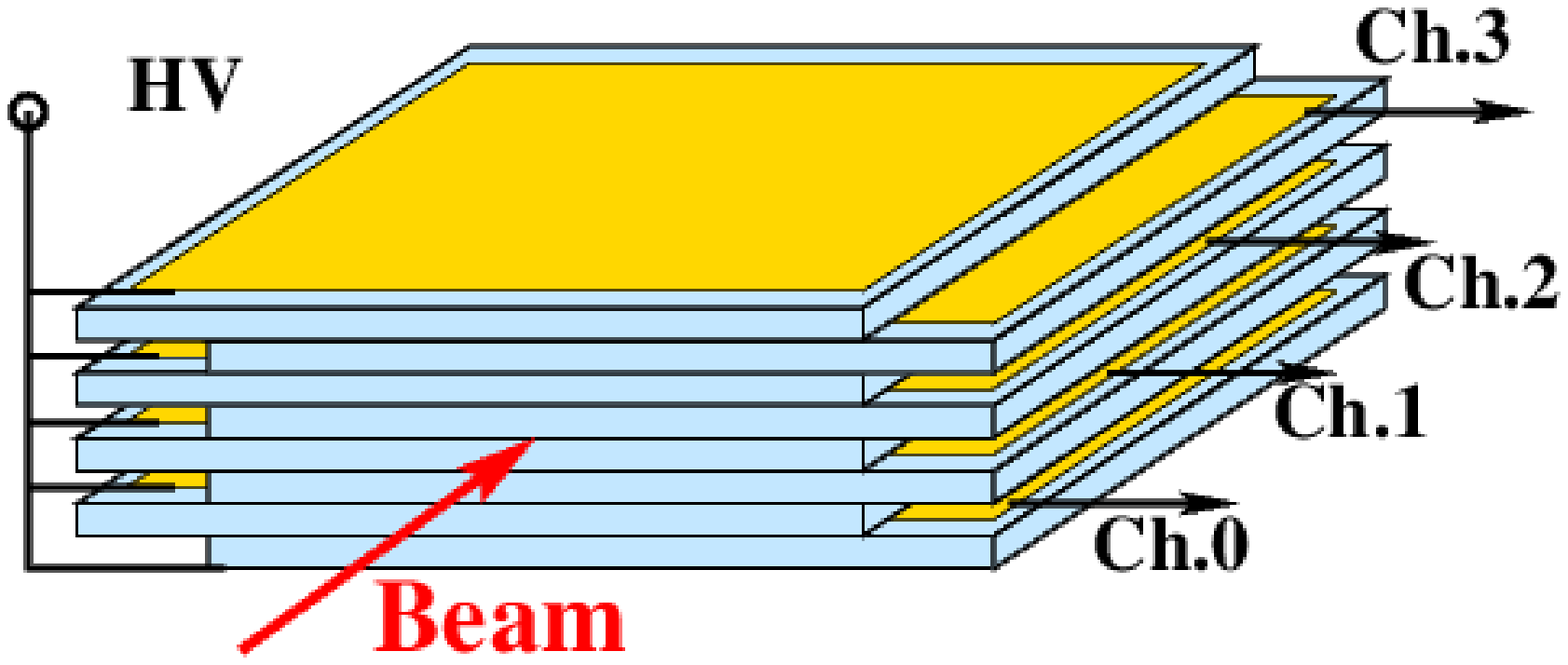}}

   \subfigure[]
  {\includegraphics[width=7.0cm]{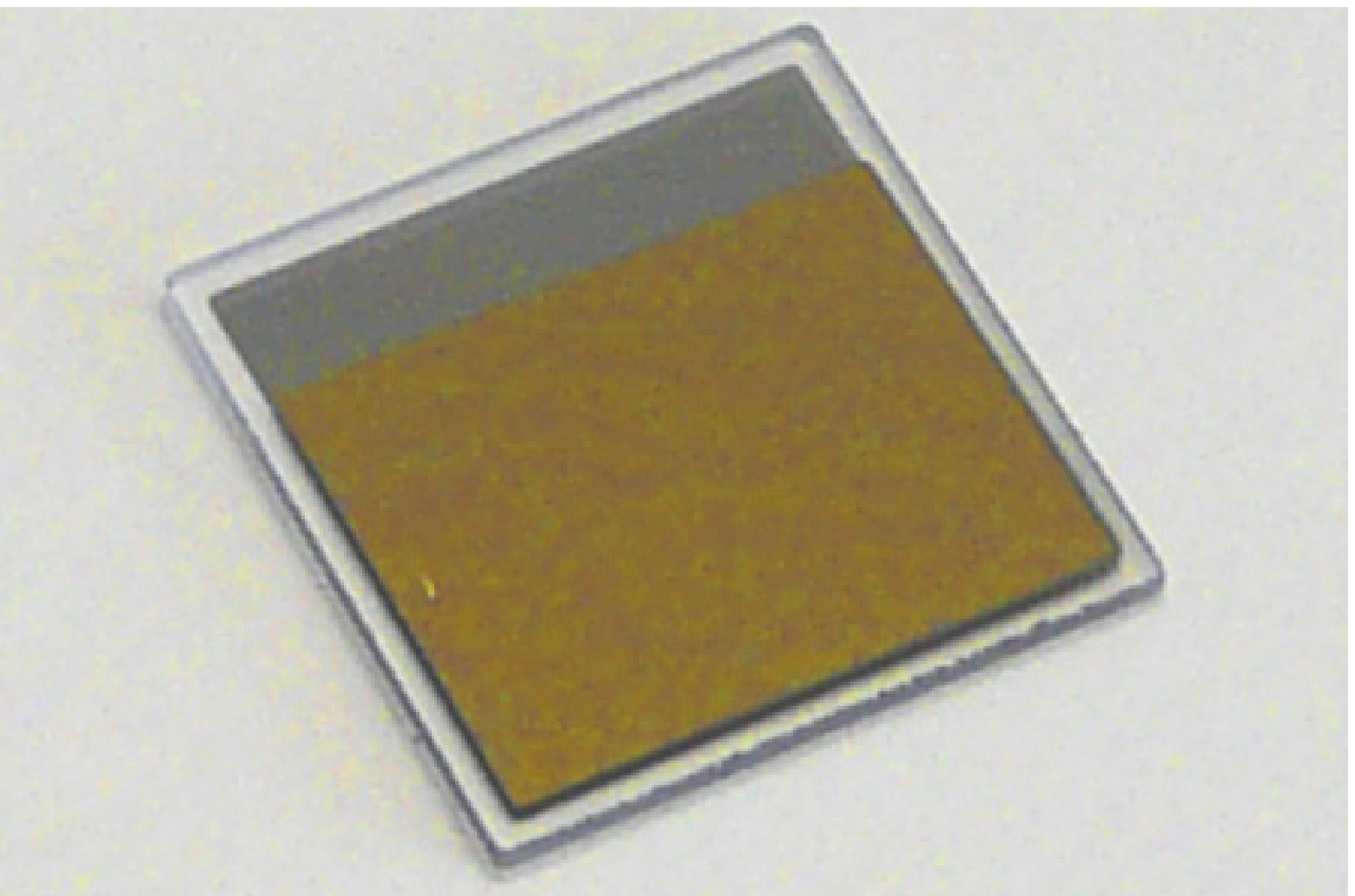}}
   \subfigure[]
  {\includegraphics[width=7.0cm]{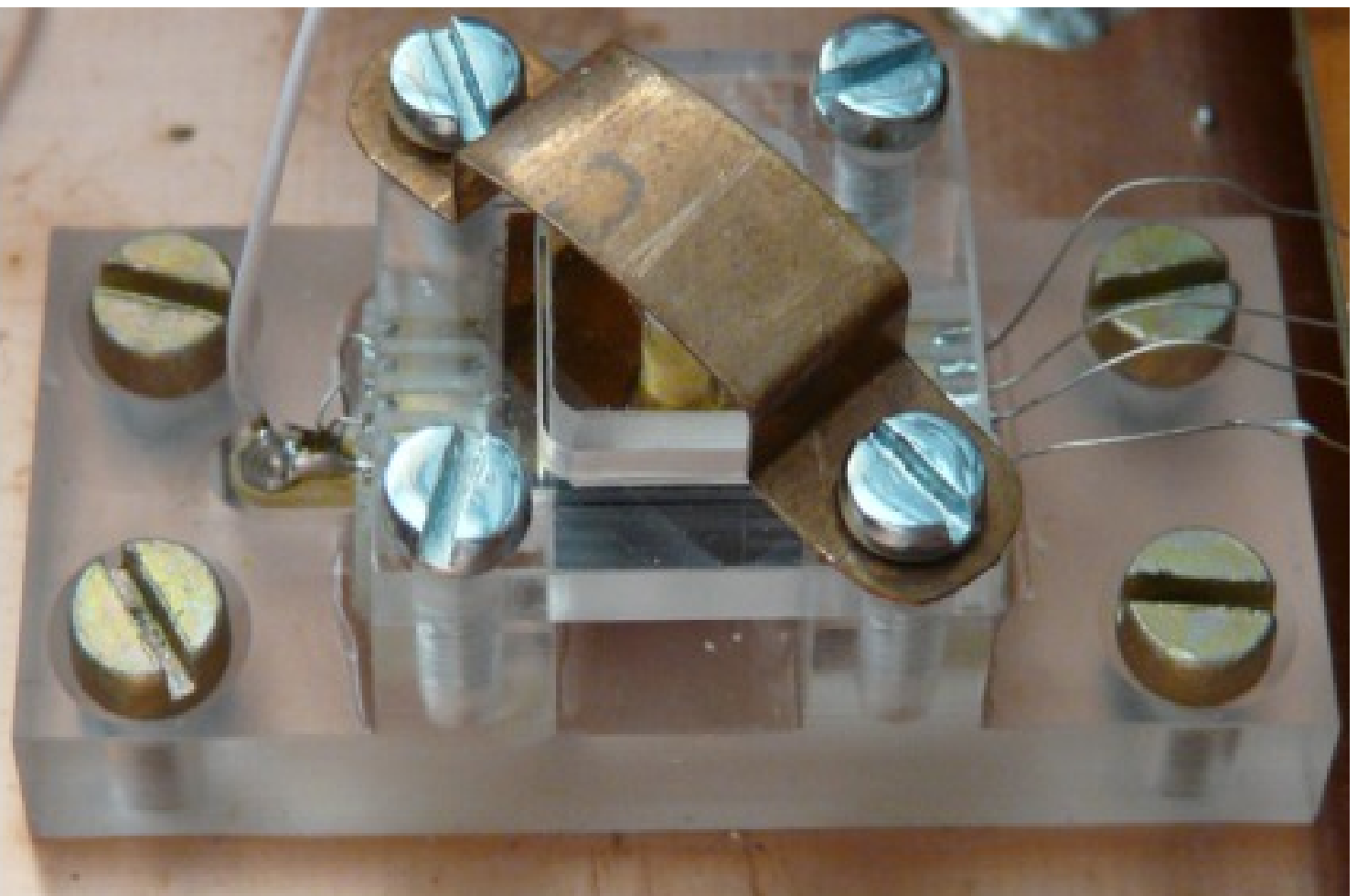}}
   \end{center}
\caption []{\label{fig:SapphireSensor}
The sapphire detector under test:
(a) -- Orientation of single sapphire sensor with respect to the beam.
(b) -- Schematic view of detector stack consisting of eight metallized sapphire sensors.
(c) -- Metallized sapphire sensor. 
(d) -- Assembled detector stack as used in the test beam.

}
\end{figure}

For a CCE of about 10\% of industrially produced sapphire~\cite{these-ignatenko}, the signal 
expected for particles crossing a plate of $500~\rm{\mu m}$ thickness perpendicular to its surface is only about $1100$~e. 
However, if the particle crosses the sapphire sensor parallel to the $10\times 10\rm{~mm^2}$ metallized surface, 
as shown in Figure ~\ref{fig:SapphireSensor}~(a), 
the signal is enhanced by a factor 20, amounting to about $22000$~e, comparable to the one in currently used solid state detectors. 
Therefore the orientation of the sapphire plates in the test beam was chosen to be parallel to the beam direction. 
In addition, this orientation leads to a direction sensitivity. Only particles crossing the sensor parallel to the surface create
the maximum signal.
To increase the effective area of the detector, eight plates were assembled together. 
To allow wire bonding connections to the high voltage and to the readout electronics, the plates were alternatively shifted to both sides. 
Each readout channel served two plates, as can be seen in Figure~\ref{fig:SapphireSensor} (b). 
Each sensor has dimensions $10\times 10\times 0.525~\rm{mm^3}$, metallized from both sides with consecutive 
layers of Al, Pt and Au of $50$~nm, $50$~nm and $200$~nm thickness, respectively. 
On one side, shown on the top plate in Figure~\ref{fig:SapphireSensor} (b), 
the metallization has a square shape of $9\times 9~\rm{mm^2}$ area. On the opposite side the metallization 
area is $9\times 7~\rm{mm^2}$ 
with $9$~mm parallel to the beam direction, as shown in Figure~\ref{fig:SapphireSensor} (c).  
 This way an accidental contact of high voltage wire bonds with the readout pad on the adjacent sensor is excluded.
The total height of the stack was $4.2$~mm with $7$~mm overlap of the metal pads, leading to a sensitive area of $29.4~\rm{mm^2}$. 
The sensors were mounted inside a plastic frame as shown in Figure~\ref{fig:SapphireSensor} (d).
The wire bonds for high voltage and readout connections are seen at the left and right side, 
respectively, in Figure~\ref{fig:SapphireSensor} (b) and (d). 
The leakage current of each pair of sensors was measured to be below $10$~pA at $1000$~V. 

\section{Test beam setup }

\begin{figure}[ht!]
  \begin{center}
  {\includegraphics[width=15.0cm]{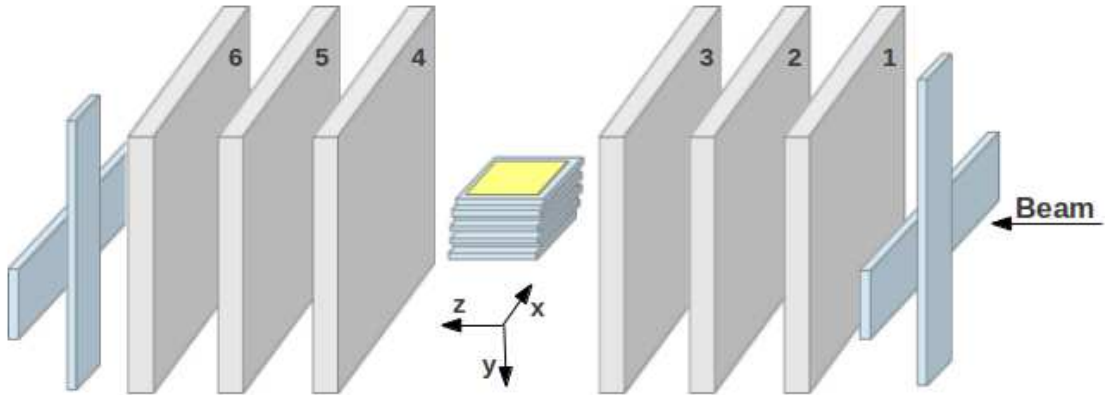}}
   \end{center}
\caption []{\label{fig:TB_setup}
Sketch of the test beam setup. The sapphire stack was mounted in the middle of the 6 planes of the EUDET telescope. 
Crosses of two scintillators upstream and downstream of the telescope were used as the trigger for the DAQ.
}
\end{figure}

The stack was mounted in the middle of six planes of the EUDET pixel telescope~\cite{EUTelescope} in the $5$~GeV 
electron beam of the DESY-II accelerator.  
Signals from sensors were amplified and shaped by charge sensitive preamplifiers A250~\cite{amptek} and 
RC-CR shapers with a peaking time of $100$~ns
and digitised by a $500$~MS/s flash ADC v1721~\cite{caen}. 

Two pairs of scintillators, shown as light blue planes in Figure~\ref{fig:TB_setup} upstream and 
downstream of the telescope, were used as trigger to readout the
telescope and the sensors.
The EUDET telescope is instrumented with Mimosa26 sensors, comprising $576\times 1152$ pixels each, 
with a pixel size of $18.4\times 18.4~\rm{\mu m^2}$. 
The telescope planes were grouped. Planes $1-3$ form the first arm, and planes $4-6$ the second arm. 
Tracks of beam electrons were reconstructed for each group separately.
From special alignment runs the width of residual distributions was measured to be below $4~\rm{\mu m}$.
From a Monte Carlo study the maximum displacement of the trajectory of a $5$~GeV electron due to multiple scattering in the stack 
was estimated to be $10~\rm{\mu m}$.

\section{Data synchronization and analysis}

For the synchronization of the EUDET telescope and the stack readout a dedicated trigger logic unit, 
TLU~\cite{EUTelescope},  was used. 
For each trigger the TLU distributed a trigger sequence to the EUDET telescope and the stack data acquisitions, 
such that a unique correspondence between
records from both readouts was ensured. 

The standard telescope analysis software~\cite{EUTelescope} was used to convert hits in the EUDET telescope into space-points 
in the user geometry with the origin of the coordinate system as shown in Figure~\ref{fig:TB_setup}. 

\begin{figure}[ht!]
  \begin{center}
  {\includegraphics[width=14.0cm]{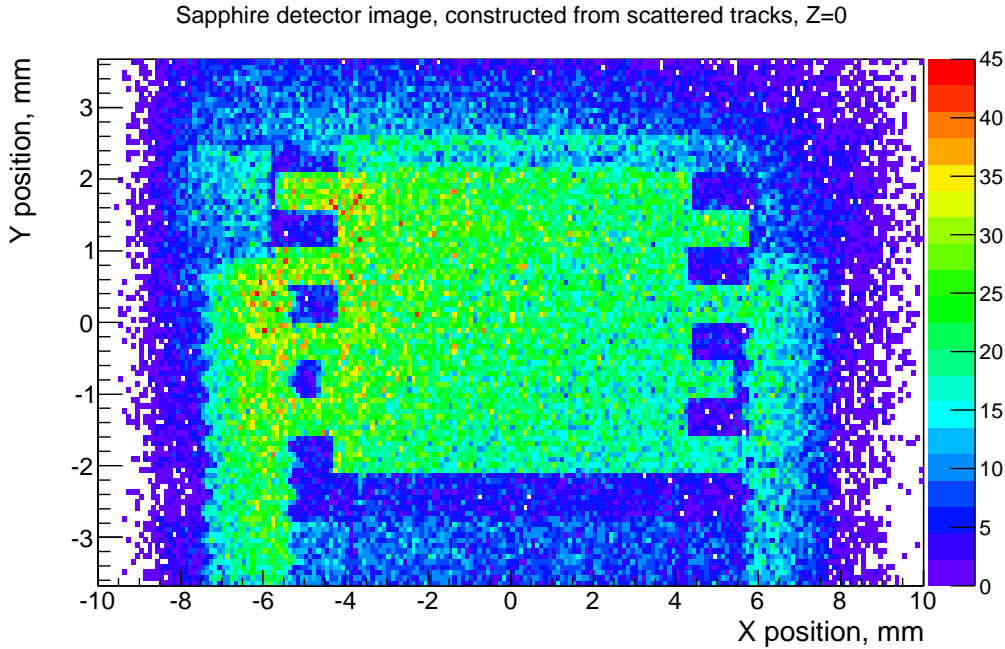}}
   \end{center}
\caption []{\label{fig:DetectorRecXY}
The image of the stack, obtained selecting tracks reconstructed in the first and second telescope arm with an angle larger than $0.5$ mrad.
Counting plates from top to bottom, the two top plates correspond to readout channel 0 and the two bottom plates to readout channel 3. 
}
\end{figure}

Events with more than one track candidate in the telescope, amounting to about $~30\%$, were rejected. 
For the remaining events the track fit was done separately for the first and second arm of the telescope. 
The two reconstructed tracks are considered to originate from the same beam electron if their distance in the $z=0$ 
plane was less than a predefined cut. 
Events matching this requirement were grouped in two sub-samples depending on the angle between the two tracks. The first sub-sample
contains events 
with an angle between the two tracks
larger than $0.5$~mrad and the second events with an angle less than $0.5$~mrad. The number of events in each sub-sample is almost the same.
Events of the first sub-sample were used   
to determine the precise position of the stack in the beam.
From the distribution of the impact points of the track of the first arm
at $z = 0$ an image of the stack in the $xy$ plane at $z=0$ is obtained, as shown in Figure~\ref{fig:DetectorRecXY}. 
From the precise position of each plate geometrical cuts were applied to select hits in each readout channel separately. 
Counting plates in Figure~\ref{fig:DetectorRecXY} from top to bottom, the top two plates one and two correspond to readout channel zero. 

\begin{figure}[ht!]
  \begin{center}
  {\includegraphics[width=12.0cm]{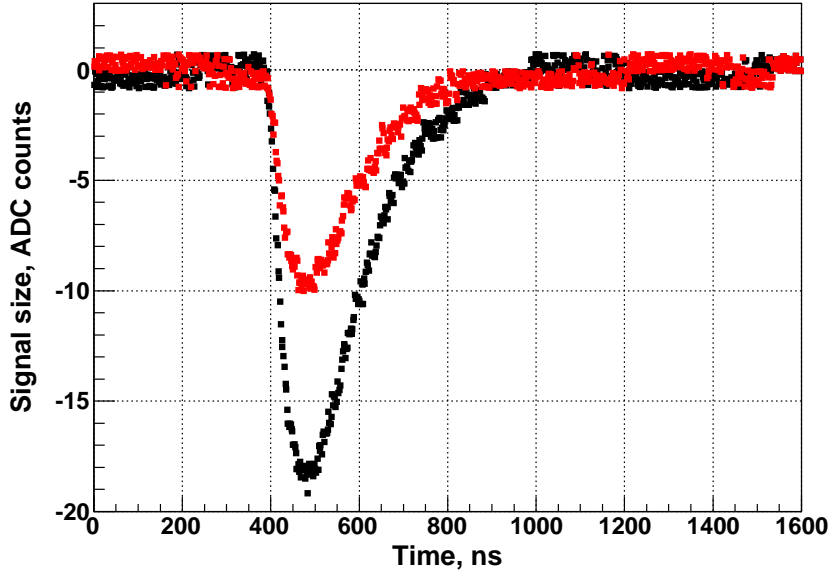}}
   \end{center}
\caption []{\label{fig:AverSignal}
The averaged signal at $550$~V in red and at $950$~V in black for events with tracks hitting plate one of the stack after 
common noise subtraction.
}
\end{figure}

Events of the second sub-sample were used for the further analysis.
For tracks of the first arm, pointing into the detector area, the $x,y$, position at $z = 0$
was determined. The fiducial area was defined using the coordinate system of Figure~\ref{fig:DetectorRecXY}
as -$4$~mm < x < $4$~mm and -$2.1$~mm < y < $2.1$~mm.
Signals from the ADC in the channel reading out the plates at the corresponding $x,y$ position were averaged over a large number of triggers. 
Events with tracks not pointing into the detector, corresponding to the blue area in Figure~\ref{fig:DetectorRecXY}, were 
used to study common mode noise. 
Correlations between the baseline values were investigated for all combination of channels using the baseline values calculated in a 
predefined time window. 
These correlations were used in the further analysis for common mode noise subtraction.
The averaged ADC output assigned to tracks not pointing into the detector was used to subtract the baseline from the averaged signal. 
As an example, the results for bias voltages of $550$~V and $950$~V for plate 1 are shown in Figure~\ref{fig:AverSignal}. Similar results
are obtained for the other plates with slightly different amplitudes as discussed in detail below. 

\section{Charge collection efficiency}

The CCE is defined as ratio of the measured to the expected signal charge. 
The signal charge is obtained by the integration of the ADC output over a $50$~ns time interval in the range [$420;470$]~ns. 
This time interval is less than the signal length as shown in Figure~\ref{fig:AverSignal}, 
but the RMS of the common mode noise in this range is low in comparison 
to the one in the tail of the signal\footnote{The rms of the common mode noise in the wave form shown in Figure~\ref{fig:AverSignal} 
is growing above 470 ns by a factor of 3 caused by the TLU. In the average waveform this is not seen, however when signals 
taken event-by-event the signal integral distributions become broader.}     . 
The mean value of the distribution of the signal charge was used for the CCE calculation at each bias voltage value.

In order to convert the mean value into a charge each channel was calibrated by injecting a known charge into the preamplifier input.
The expected amount of generated electron-hole pairs is estimated from the mean value of the ionisation energy loss inside the sensor,
obtained from a GEANT~\cite{GEANT} simulation, and the energy needed to create an electron-hole pair, as given 
in Table~\ref{tab:table1}. This quantity is estimated using the extrapolation proposed in Ref.~\cite{N_El_Holes}, and corrected 
for wide band-gap semiconductors~\cite{NEH_wide_BG}.

\begin{figure}[ht!]
  \begin{center}
  {\includegraphics[width=11.0cm]{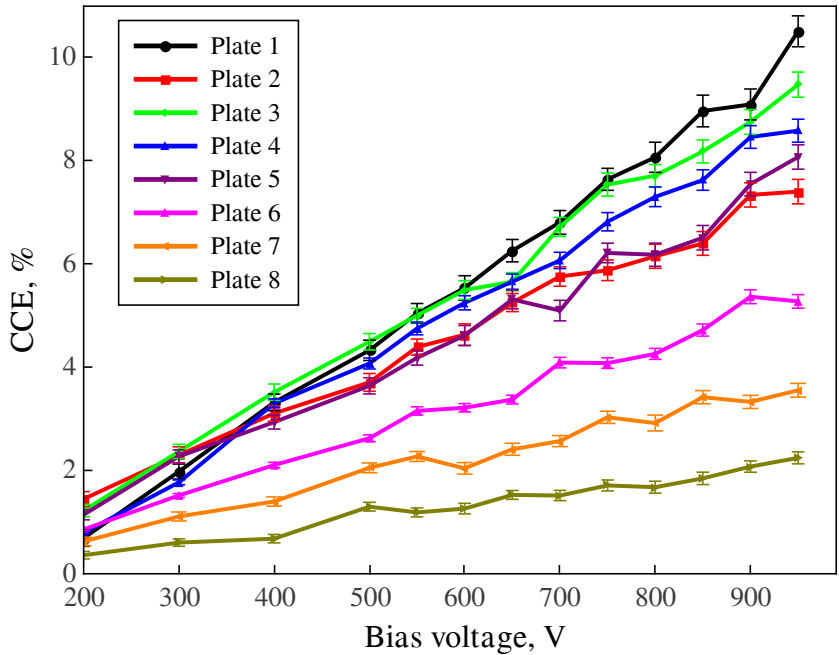}} %CCE.eps
   \end{center}
\caption []{\label{fig:CCESaph}
Measurement of the mean charge collection efficiency for eight sapphire plates as a function of the bias voltage. 
}
\end{figure}

The measured CCE is shown in Figure~\ref{fig:CCESaph} as a function of the bias voltage for all plates of the stack. 
For each voltage value a statistics of $100000$ triggers was used. To exclude signals with reduced amplitudes expected
at the edges of the metal pads due to distortions of the electric field the fiducial area was reduced 
in $x$ to $-3$~mm < $x$ < $3$~mm.
An almost linear rise of the CCE is observed, reaching at $950$~V e.g. for plane $1$ a value of $10.5$\%. 
The values of the CCE obtained for all plates at a voltage of $950$~V are listed in Table~\ref{table:CCE}. 
The statistical error is obtained as the standard deviation of the mean value. The systematic uncertainty is the uncertainty due to the calibration and the 
uncertainty due to transition region between adjacent plates in the $y$ coordinate, added in quadrature.
The measured CCE varies from sample-to-sample reflecting variation of the substrate quality. 
As can be seen, $5$ out of the $8$ sensor plates have a relatively high and similar CCE of about $7-10\%$, while three other plates have lower and different CCE values. 
A quantity <CCE> was defined as the averaged value of $12$ CCE measurements for each plate in $500~\rm{\mu m}$ steps of the $x$ coordinate.
Its values are given together with the RMS in Table~\ref{table:CCE}.
The CCE obtained from the voltage scan at $950$~V and the average <CCE> are in agreement within the uncertainties of the measurements.

\begin{table}
\begin{center}
\caption{The measured CCE at the highest applied bias voltage of $950$~V with statistical and systematic uncertainties. 
Also given are the quantities <CCE> and RMS obtained from averaging CCE measurements in 500~$\rm{\mu m}$ steps in the $x$ coordinate.}
\label{table:CCE}
\begin{tabular}{|l|c|c|c|c|c|c|c|c|}
    \hline
    Plate number  &   1   &   2    &   3   &   4   &   5    &   6   &   7   &   8   \\
    \hline
    CCE, \%       & 10.5  &   7.4  &  9.5  &  8.6  &   8.1  &  5.3  &  3.6  &  2.2   \\
    \hline
    Stat. error   & 0.3   &   0.2  &  0.2  &  0.2  &   0.2  &  0.1  &  0.1  &  0.1   \\
    \hline
%    Syst. error   & 0.6   &   0.4  &  0.5  &  0.5  &   0.4  &  0.3  &  0.2  &   0.1  \\ % 5% of CCE
    Syst. error   & 0.2   &   0.2  &  0.2  &  0.2  &   0.2  &  0.1  &  0.1  &   0.1  \\ % 2% of CCE
    \hline \hline
    <CCE>, \%     & 9.9   &   7.2  &  9.0  &  8.5  &   7.5  &  5.3  &  3.7  &  2.1   \\
    \hline
    RMS           & 1.7   &   0.9  &  0.9  &  0.8  &   0.6  &  0.5  &  0.9  &  0.7   \\
    \hline

\end{tabular}
\end{center}
\end{table}

\section{Theoretical model for the charge collection efficiency of sapphire sensors}

A linear model was developed to describe the CCE as a function of the local $y$ coordinate inside a plate. 

Charged particles cross the sensor and ionize the atoms along their path through the sensor of thickness $d$. $N_0$ electron-hole pairs are produced.
Some charge carriers will recombine immediately. A fraction of both types of charge carriers, called $f_d$,
start to drift to the corresponding electrodes when an external electric field is applied. During the drift charge carriers may be trapped.
Trapped charges may after some time recombine, or, to a certain fraction, thermally released. In the latter case they contribute to the signal current. 
If the occupation of traps is small and detrapping time
is significantly shorter than the duration of the measurement, the density of trapped charges in steady state will be proportional to the flux of drifting
charge carriers. The space charge due to trapped charges generates an internal electric field, 
called polarization field, with the direction opposite to the externally applied field. 
Assuming that the space charge density will be a linear function of the local $y$, the resulting electric field has a parabolic shape,
in the simplest case of a non-charged crystal $ E(y) = {A (y - \frac{d}{2}) ^2} + B $, where $A$ and $B$ are parameters.
The integral of the electric field over the full sensor thickness $d$ is equal to the bias voltage.

To estimate the signal size electrons and holes will be considered separately, as they may contribute to the resulting signal differently. 
The drift velocity $ v_{e,h}$ is assumed to be directly proportional to the electric field strength, $ v_{e,h} = \mu_{e,h} E(y)$ , 
where $\mu_{e,h}$ is the mobility for electrons and holes, respectively. 

\begin{table}
\begin{center}
\caption{Parameters obtained from the fit of the CCE measured as a function of the local $y$ using eqn. 8.1 at a voltage of $950$ V.}
\label{table:fit_param}
\begin{tabular}{|c|c|c|c|c|c|}
    \hline
    Plate number &   B, $V/\mu m$    &  $f_d$, \%      & $\mu \tau (e),~\mu m^2/V$ & $\mu \tau (h),~\mu m^2/V$ & $\chi^2$    \\
    \hline
    1            & 1.327 $\pm$ 0.012 &  52.9 $\pm$ 0.5 &       79.1 $\pm$ 1.1      &    4.2 $\pm$ 0.3          &  19         \\
    \hline
    2            & 1.255 $\pm$ 0.011 &  47.1 $\pm$ 0.5 &       59.5 $\pm$ 0.9      &    6.2 $\pm$ 0.3          &  41         \\
    \hline
    3            & 1.307 $\pm$ 0.010 &  53.3 $\pm$ 0.5 &       64.9 $\pm$ 0.9      &    6.4 $\pm$ 0.2          &  27         \\
    \hline
    4            & 1.287 $\pm$ 0.011 &  48.1 $\pm$ 0.5 &       74.6 $\pm$ 1.0      &    3.3 $\pm$ 0.3          &  27         \\
    \hline
    5            & 1.421 $\pm$ 0.010 &  47.1 $\pm$ 0.7 &       62.9 $\pm$ 1.0      &    3.2 $\pm$ 0.4          &  16         \\
    \hline
    6            & 1.342 $\pm$ 0.013 &  43.5 $\pm$ 1.3 &       39.4 $\pm$ 1.2      &    5.1 $\pm$ 0.4          &  42         \\
    \hline
    7            & 1.484 $\pm$ 0.010 &  50.1 $\pm$ 1.2 &       22.0 $\pm$ 0.8      &    3.7 $\pm$ 0.4          &  19         \\
    \hline
    8            & 1.330 $\pm$ 0.010 &  40.7 $\pm$ 1.7 &       15.1 $\pm$ 0.5      &    3.2 $\pm$ 0.4          &  33         \\
    \hline

\end{tabular}
\end{center}
\end{table}

\begin{figure}[ht!]
  \begin{center}
   {\includegraphics[width=14.0cm]{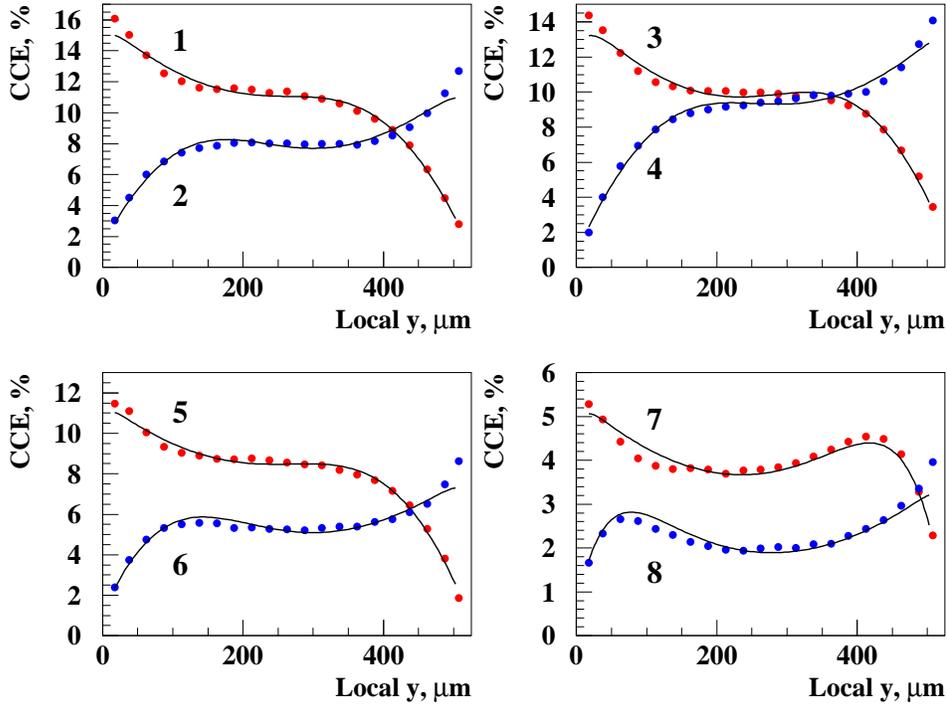}}

%   {\includegraphics[width=12.0cm]{plate58_h.eps}}
   \end{center}
\caption []{\label{fig:fit_plates}
The CCE measured as a function of the local $y$ coordinate inside a plate in slices of 25~$\rm{\mu m}$ for all plates of the sapphire stack. 
Blue dots are for the electric field in the direction of y and red dots for the opposite field direction. The lines are the result
of a fit including both electron and hole drift. The fit parameters are given in Table~\ref{table:fit_param}.
}
\end{figure}

The charge carrier lifetime $\tau_{e,h}$ is assumed to be constant. Then the number of carriers at time $t$ is $N_{e,h}(t) = f_d \cdot N_0 e ^{-t/\tau_{e,h}}$. 
According to Ramo's theorem the differential contribution to the observed signal is $dQ=\frac{e}{d} N(t) v_{e,h} dt$. Substituting $dt$ by
$dt = \frac{dy}{\mu_{e,h}E(y)}$ and integrating from the carrier generation point $y_0$ to the electrode surface, the signal charge is parametrised as:
\begin{equation}
Q(y_0)= {\frac{e \cdot f_d \cdot N_0}{d}} {\mathrm{e} ^{{\frac{arctan\left((y_0 - {\frac{d}{2}}) \sqrt{\frac{A}{B}}\right)}{\mu \tau \sqrt AB}}} 
\int_{y_0}^d \mathrm{e}^{-{\frac{arctan\left((y - {\frac{d}{2}}) \sqrt{\frac{A}{B}}\right)}{\mu \tau \sqrt AB}}}\mathrm{d}y}. 
\label{eq:charge}
\end{equation}
The quantity $B$ is the electric field strength at the plane in the middle of the plate, $y = d/2$, and $\mu \tau$  is the drift path length of the electrons or 
holes in the electric field of unit strength. %% It is worth to emphasize, that the collected charge depends on the product of $\mu$ and $\tau$.
The ratio $Q/N_0\times \rm{e}$ is then the fraction of the charge carriers contribution to the observed CCE.
In case of $\mu \tau E \ll d$ the proportionality between the charge carrier drift path and electric field strength leads to the linear dependence
of the CCE on the detector bias voltage, in agreement with the measurement shown in Figure~\ref{fig:CCESaph}.

Figure~\ref{fig:fit_plates} shows the CCE as a function of the local y coordinate for a voltage of $950$~V, measured in $25~\rm{\mu m}$ slices, 
together with 
a fit using equation~(\ref{eq:charge}). 
The electric field has opposite direction for adjacent plates. For example, $y = 0~\mu m$ of plate 1 and $y = 525~\rm{\mu m}$ of plate 2 correspond to the
same readout electrode. In plates 1, 3, 5 and 7 the electric field is directed from $y =  525~\rm{\mu m}$ to $y = 0~\rm{\mu m}$ 
and the CCE is shown in red dots. 
In plates 2, 4, 6 and 8 the field direction is opposite, and the CCE is shown in blue dots. 
The parameters of the fit are listed in the Table~\ref{table:fit_param}. 
As can be seen, the drift length of electrons is in most of the cases 
more than 10 times larger than the drift length of the holes at roughly the same field strength. 
This result is consistent with low hole mobility predicted in Ref.~\cite{hole_mass} and confirms the dominant contribution of electrons for the charge
transport in sapphire~\cite{hall_mobility,phot_cond}. 

\section{Conclusions}
The paper presents results of the performance of a multi-channel sapphire stack, designed for single particle detection, in a $5$ GeV electron beam. 
The CCE shows a linear dependence on the bias voltage reaching up to 10\% at $950$~V. A measurement of the CCE as a function of the local $y$ 
coordinate through the thickness of the plates shows a pronounced dependence on $y$. The measurement can be explained by a linear model pointing to
a dominant contribution of electron drift to the signal charge and to the presence of a polarisation field inside the bulk of the sensor. In addition,  
a fraction of charge carriers of about 50~\% recombines immediately after creation.

\section{Acknowledgments}
This work was supported by the Commission of the European Communities under the 7th Framework Program AIDA, contract no. 261015.
The Tel Aviv University and the Brandenburgische Technische Universit{\"a}t Cottbus are supported by the German-Israel Foundation (GIF).
We are grateful to the crew and the management of the S-DALINAC accelerator at the Technical
University of Darmstadt for their great support during the measurements at the electron beams.
The measurements leading to these results have been performed at the
Test Beam Facility at DESY Hamburg (Germany), a member of the Helmholtz Association (HGF).
We would also like to thank the Target Lab of the GSI Helmholtzzentrum f{\"u}r Schwerionenforschung for the metallization of the sensors.

\end{document}